# Quasi-ballistic phonon transport effects on the determination of the mean-free path accumulation function for the effective thermal conductivity

Ashok T. Ramu[†] and John E. Bowers

University of California Santa Barbara

[†]Corresponding author: ashok.ramu@gmail.com

The mean-free path (MFP) accumulation function for the effective thermal conductivity, introduced by Dames and Chen is a compact, universal and highly useful summary of the effect of ballistic thermal transport on the effective thermal conductivity measured by various experiments. The frequency domain thermoreflectance (FDTR) experiment is especially well suited to its determination. Extraction of the accumulation function from this experiment uses the thermal penetration depth (TPD) of phonons at each frequency as a cut-off for classifying phonons as ballistic or diffusive at that frequency. In this paper, we show that using the TPD as a cut-off is arbitrary and prone to serious error. We report on a new technique to deduce the MFP accumulation function from the FDTR experiment by numerical solution of an enhanced Fourier law.

## 1. Introduction

The mean-free path accumulation function (MFPAF) introduced by Dames and Chen [1] is a powerful tool in studying ballistic phonon transport. The MFPAF at a given mean-free path Λ is defined as the effective thermal conductivity ($\kappa_{eff}$) contributed by all phonons with mean-free paths less than or equal to Λ. The utility of the MFPAF lies in that it explains within a unified framework[2] diverse experiments, like the transient gratings[3], time-domain thermoreflectance (TDTR)[4][5] and frequency domain thermoreflectance (FDTR)[6] experiments, that probe heat transport on length scales comparable to the phonon mean-free path.

For bulk materials, measurements of the MFPAF have been conducted for crystalline[7] and amorphous[8] materials. For nanostructured materials, Yang and Dames [9] have given a relationship to the MFPAF of bulk materials. The MFPAF of nanostructured materials has been determined by measuring length-dependent conductivity in nanowires[10]. The MFPAF has been applied to the determination of thermal properties of nanostructured materials from bulk properties [11]. The concept of MFPAF has been applied to the study of thermal interfaces as well [12].

The thermoreflectance class of measurements was invented by Paddock and Easley [13]. The FDTR experiment is especially well suited to the determination of the MFPAF. In the FDTR experiment, a sinusoidally modulated "pump" laser beam impinges on the sample, and its modulation frequency is varied across a wide range. The resulting surface temperature is monitored by means of a weak "probe" beam reflected by a thin metal transducer layer that coats the sample, and whose reflectivity is sensitive to the temperature fluctuations caused by the pump. As the modulation frequency ν increases, the thermal penetration depth (TPD), which is the depth at which the temperature oscillation amplitude equals 1/e times the surface value, falls as $\sqrt{1/\nu}$ in the weakly quasi-ballistic limit of phonon transport [20]. This limit is valid only when the modulation frequency is much lesser than the inverse lifetime of the dominant heat carriers. The TPD is a measure of how deeply the thermal wave penetrates the sample. Phonons with mean-free paths much greater than the TPD are supposed to fail to equilibrate with the lattice within a depth equal to the TPD, and are therefore presumed lost to the measurement. Thus the "effective" thermal conductivity ($\kappa_{eff}$) decreases from its bulk value as the modulation frequency increases [6].

Throughout this paper we focus on silicon bulk substrates with a fixed and somewhat arbitrary, but reasonable boundary thermal conductance of 200 MW/m$^2$ -K between the metal transducer and the silicon. This value is in concord with the fitting of Regner *et al.* to the high-frequency (~100 MHz) FDTR data [6]. The results of this paper, especially the newly extracted MFPAF (Fig. 6), are particularly robust to boundary thermal conductances between 50-1000 MW/m$^2$-K. Silicon has a broad MFPAF with phonons of almost 3 orders of magnitudes of MFPs contributing non-trivially to thermal transport [14].

In this paper, we show that the choice of the TPD as the cutoff between ballistic and diffusive modes is intrinsically arbitrary and leads to significant error in the extracted MFPAF, especially in the low MFP regime. Arbitrariness may be due to two reasons: the choice of the 1/e depth as opposed to, say, the 1/e$^2$ depth is one source. This is just a matter of choice and we will not focus on this here.

The other and more serious source of error lies in that, as will be shown below, phonons with MFP equal to the TPD contribute significantly (roughly 42% at 88 MHz in silicon at 300 K) to the net heat-flux. Clearly, ignoring phonon modes with this MFP as fully ballistic (instead of quasi-ballistic) leads to erroneous conclusions about the MFPAF.

This paper is organized as follows: First, we briefly introduce the theoretical model we use to derive our results and state its main features. Next, we describe the methodology we use to derive the cut-off for the MFP of quasi-ballistic phonons. In doing so, we introduce expressly for the first time the concept of a frequency-dependent MFP accumulation function. We follow by examining in detail the effect of cutting off the diffuse modes at the TPD (hereafter called the TPD-cutoff model). We then apply our theoretical model and show that it yields cutoffs far from the TPD, giving us a very different MFPAF. We conclude by summarizing our findings.

## 2. Theoretical model

Our theoretical analysis is based on the enhanced Fourier law (EFL) [15] introduced by one of the authors. The basis of our model of thermal transport is the "two-fluid" assumption[16], wherein the phonon spectrum is divided into two parts- one a high-heat-capacity, high-frequency (HF) part that is in quasi-thermal equilibrium with a well-defined local temperature, and the other, a low-frequency (LF), low-heat-capacity part that is farther out of equilibrium. The LF modes do not interact with each other due to the small phase-space for such scattering[17], but can exchange energy with the HF modes. This model accounts for the effect of low-frequency phonon modes of long mean-free path, that propagate concomitantly to the dominant high-frequency modes. The cutoff for classification of modes as HF and LF will be the subject of much discussion in this paper.

The theory of the EFL is based on spherical harmonic expansions of the phonon distribution functions, wherein the high-frequency mode distribution function is truncated at the first order in the expansion, while the low-frequency mode distribution function, which is farther out of thermal equilibrium, is truncated at the second order. This procedure has the advantage that successive terms of the spherical harmonic expansion may be directly related to quantities of physical interest like the energy-density and heat-flux. The EFL, which may be viewed as a non-local refinement of the Fourier law, is written as

$$\frac{\partial q}{\partial x} = -C_v \frac{\partial T}{\partial t} + S^{HF}(x,t) \qquad (1)$$

$$q = \frac{3}{5}(\Lambda^{LF})^2 \frac{\partial^2 q}{\partial x^2} + \frac{3}{5}\kappa^{HF}(\Lambda^{LF})^2 \frac{\partial^3 T}{\partial x^3} - \kappa \frac{\partial T}{\partial x} \qquad (2)$$

Here, the net heat-flux $q= q^{LF}+q^{HF}$, $q^{LF}=$ the LF-mode contribution to the heat-flux, $q^{HF}=$ the HF-mode contribution to the heat-flux; $C_v$ is the volumetric heat capacity; $S^{HF}(x,t)=$ external heat source term; $T$ = local temperature of HF modes; $\kappa$ is the net bulk thermal conductivity; $\kappa = \kappa^{LF} + \kappa^{HF}$, $\kappa^{HF}=$ the contribution of HF modes to the bulk thermal conductivity, $\kappa^{LF}=$ the contribution of LF modes to the bulk thermal conductivity; and $\Lambda^{LF}=$ the MFP of the LF modes $=v\tau$ where $v$ is the group-velocity magnitude of all LF modes and $\tau=$ LF mode lifetime. We assume that each and every LF mode has the same lifetime $\tau$, as well as the same group-velocity magnitude $v$. The frequency of an LF mode of wave-vector $\boldsymbol{k}$, denoted by $\omega(k)$ is permitted to vary with $k$. We also assume isotropic phonon dispersion.

For future reference, we state equations for the LF mode and HF mode heat-fluxes $q^{HF}$ and $q^{LF}$ separately:

$$q^{LF} = \frac{3}{5}(\Lambda^{LF})^2 \frac{\partial^2 q^{LF}}{\partial x^2} - \kappa^{LF} \frac{\partial T}{\partial x} \qquad (3)$$

$$q^{HF} = -\kappa^{HF} \frac{\partial T}{\partial x} \qquad (4)$$

It has been shown in an earlier article that truncation of the LF mode distribution function at the second order in the spherical harmonic approximation is a good approximation[19] and that an excellent match with the transient gratings experiment in silicon at 300 K is obtained with an LF phonon MFP of Si (400 nm) that agrees well with Ref. [20], thus validating our model. These equations are derived from the model assumptions in Appendix A.

We adapt the EFL to the conditions of the FDTR experiment by assigning a time-dependence of the form $e^{i\omega t}$ to all variables, where ω is the modulation (angular) frequency, and then by working entirely in the Fourier domain. In that case, for the substrate, where there is no heat source, substituting Eq. (1) into the derivative (with respect to *x*) of Eq. (2), we obtain a fourth-order linear homogeneous differential equation for (complex) *T*. Its four characteristic roots were found numerically using MATLAB®. Assuming an infinite substrate with all heat-fluxes and temperatures set to 0 at the bottom, we may drop the two exponentially growing solutions. Thus there remain two constants of integration.

These constants are determined using continuity of heat-flux between the transducer and substrate: we set $q^{LF}=0$ and $q^{HF} = Q$ (Eqs. (3), (4)) at the silicon surface, where $Q$ is the net heat-flux emanating from the transducer; thus the surface temperature of silicon in the absence of an interface may be found. The thermal interface is accounted for by a temperature drop *ΔT=Q/G* where *Q* is the heat-flux and *G* is the boundary thermal conductance. Finally $Q$ is related to the input from the "pump" laser by solving the Fourier law in the tranducer film, with heat-capacity 1.6X10$^6$ W/m$^3$-K, thermal conductivity 100 W/m-K and thickness 100 nm as appropriate for an Au transducer [6]. The phase of the temperature is of interest since it is robust to fluctuations in laser power[21], and is calculated as the arctangent of the ratio of the imaginary to the real part of the complex temperature.

The only subtlety here is that we assume all of the heat emitted by the "pump" laser into the transducer transfers exclusively to the HF modes of silicon. This is because heat from the laser scatters into final states in silicon in proportion to their density-of-state, according to the Fermi golden rule [22]. Since the density-of-state for acoustic phonons varies approximately as the square of the frequency[24], HF modes are excited much more efficiently than LF modes by external heat sources.

## 3. Methodology

We consider in Fig. 1 the transducer phase as a function of logarithmic frequency. This function is so designed as to replicate, for hypothetically large "pump" laser spot diameters (≥ 100 micrometer), the Fig. 4 of Regner *et al.* [6] for the MFPAF of silicon at room temperature, using the TPD-cutoff model. In other words, the phase of Fig. 1 at each frequency is fitted to the usual Fourier law with an effective conductivity $\kappa_{eff}$, and the corresponding frequency $\omega$ is translated into an MFP by setting the MFP equal to the TPD, $\sqrt{2\kappa_{eff}/\omega C_v}$. Here $C_v$ is the volumetric heat capacity of silicon. We thus recover the MFPAF, Fig. 4 of Regner *et al.*, from our assumed phase data. We choose not to use the experimental phase data of Regner *et al.* because their small spot size (~ 3.4 microns) would cause significant radial ballistic effects[22] which would complicate the subsequent analysis. Fig. 4 of Ref. [6] is replotted in Fig. 6 (red solid curve).

Our main point in this paper is that when we analyze the same phase data (Fig. 1) using the enhanced Fourier law [15] instead, we recover an accumulation function very different from Fig. 4 of Regner *et al.* (Ref. [6]). To this end, at each frequency, we find the MFP $\Lambda^{LF}$ of the LF modes by iteratively varying it, using the MATLAB® routine "*fsolve*" until the heat-flux in the HF modes equals the heat-flux in the LF modes in magnitude (this criterion is discussed later). The other parameters are determined as follows: in Eqs. (1) and (2), at each frequency, we take $\kappa^{HF}$ as $\kappa_{eff}$ at that frequency, $\kappa_{eff}$ being as before the effective conducitivity according to the usual Fourier law. This is from the definition of the MFPAF and means that we leave *y*-coordinates of the MFPAF unchanged. Also, $\kappa^{LF} = (\kappa_{bulk} - \kappa_{eff})$ where $\kappa_{bulk}$ is the bulk conductivity of silicon, constant at 143 W/m-K. This is from the definition of $\kappa^{LF}$ in the EFL. The laser power input into the transducer is assumed to be $10^6/t$ W/m$^3$ where *t* is the transducer thickness in m.

In order to differentiate our methodology from that of the TPD-cutoff model, we introduce here the concept of a "modulation frequency-dependent MFP accumulation function". We note that at each frequency, the TPD-cutoff model implicitly uses this concept; its frequency dependent MFPAF consists of simply rejecting the contributions of all modes with MFP ≥ TPD. Our model of the frequency dependent MFPAF whereas consists of a cut-off MFP $\Lambda^{LF}$, not pre-set at any value, but solved for from the detailed thermal properties of the substrate. We assume that modes with MFP equal to the cut-off MFP contribute quasi-ballistically and constitute the $q^{LF}$ of Eq. (3). Fig. 2 shows a schematic of the difference between the two models.

While assigning the same MFP ($\Lambda^{LF}$) to all modes above the cutoff might seem like an oversimplification considering the broad MFP distribution of silicon, we note that for a specific frequency of modulation, most of the phonons with MFPs above the cut-off $\Lambda^{LF}$ are lost to the measurement, and hence their contribution to the bulk conductivity becomes irrelevant. At any rate, our model is superior to the TPD-cutoff model of Koh and Cahill [4], that entirely ignores quasi-ballistic conduction by phonons with MFPs at and above the TPD.

We set the criterion for the cut-off as the mean-free path at which the HF-mode heat-flux contribution (magnitude of $q^{HF}$ of Eq. (4)) drops to the point where it equals the LF-mode contribution (magnitude of $q^{LF}$ of Eq. (3)). At all frequencies, it is observed that this criterion ensures that the LF-mode heat-flux from phonons of MFP equal to this cut-off, and therefore from phonons of MFP greater than this cut-off, is negligible compared to the net heat-flux.

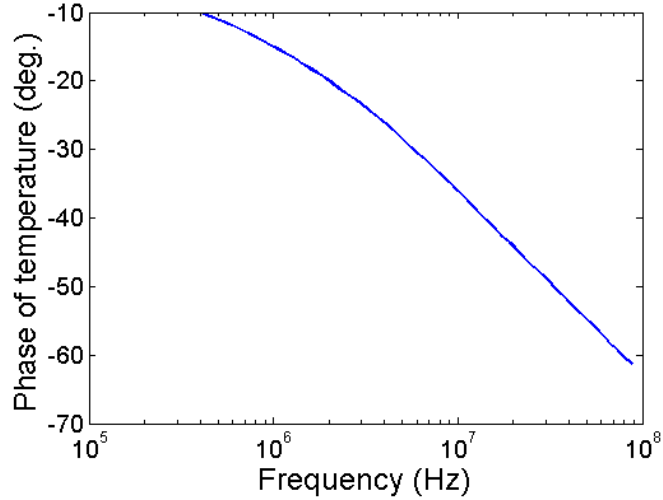

Fig. 1: Phase of the transducer temperature oscillation as a function of (linear) frequency. The phase function is designed to replicate the MFP data of Fig. 4 of Regner *et al.* [6], shown in this paper as the red solid curve of Fig. 6.

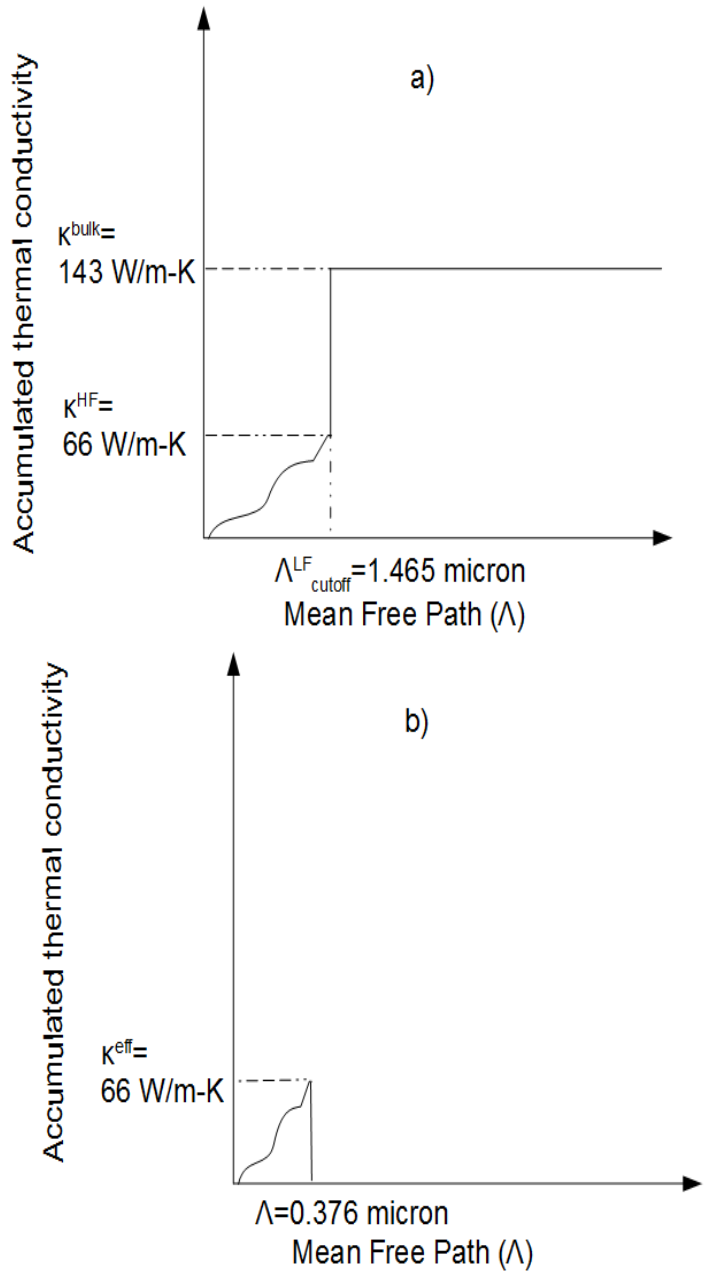

Fig. 2: A schematic representation of the modulation frequency-dependent MFPAF at a modulation frequency of 88 MHz for (a) our EFL-based model, and (b) the TPD-cutoff model. Our model accounts for quasi-ballistic transport above the MFP cutoffs (abscissae of the vertical lines).

## 4. Results and discussion

We plot in Fig. 3 the heat-fluxes vs depth under the TPD-cutoff model. It is seen that LF-channel phonons with an MFP of 376 nm contribute 42% of the net heat-flux at 88 MHz, when measured at a depth of 376 nm (equal to the TPD at that frequency) from the surface. Also, the LF-mode heat-flux at that depth is about 25% of the surface heat-flux ($8 \times 10^5$ W/m$^2$). In Fig. 4, using the EFL-based model of this work, with the cut-off $\Lambda^{LF}$ set at 1.47 micrometers this contribution is reduced to 6% of the net heat flux at the same depth (376 nm). Now we may safely say that phonons with MFP greater than the cutoff contribute negligibly to the measurement.

Fig. 5 shows the cut-off MFP vs frequency for both this work and the TPD-cutoff model. It is seen that our cut-off is dramatically larger than the cut-off based on the TPD. Also, our model is not a simple scaling of the TPD-cutoff by a constant factor. At low frequencies, our cut-off somewhat resembles 8 times the TPD, but even this connection breaks down at higher frequencies by being a factor of 2 too great. We further note that the quasi-ballistic heat-flux $q^{LF}$ is rather insensitive to the mean-free path cut-off $\Lambda^{LF}$ at low-frequencies. Thus it may be seen that a slightly different cut-off criterion might give quite different values of the MFP cut-off. This reflects, in our opinion, a fundamental inability to determine the contribution of high MFP phonons to the accumulation function from FDTR measurements. This is because they carry too little heat, e.g. 5% of the bulk value at our lowest frequency (0.4 MHz). This low value is in turn because of the low heat-capacity of these modes [24].

Fig. 6 compares the MFPAF derived from our model to that from the TPD-cutoff model. Our model clearly shifts the MFPAF derived from the TPD cut-off model to the right. This is to be expected, since quasi-ballistic phonon heat-fluxes which decay slower than diffusive fluxes are accounted for in our model. We note here that unlike Regner *et al.*[6] whose MFPAF extraction is based on the TPD model, our MFPAF is inconsistent with *ab-initio* calculations of Esfarjani *et al.* [14]. Further exploration of this discrepancy is strongly indicated. It must be emphasized that Fig. 6 should not be viewed as the "correct" and "wrong" MFPAF for silicon – it simply indicates that the MFPAF as given by the TPD-cutoff model is inconsistent with the experiment from which it is extracted.

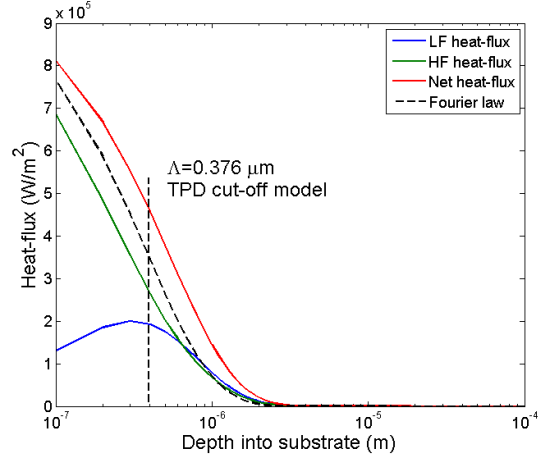

Fig. 3: Phonons with MFPs above the TPD contribute as much as 42% to the net heat-flux at 88 MHz. Their neglect results in serious error in the MFPAF. Dashed line shows "effective" Fourier law result, which is less than the EFL net heat-flux because it neglects quasi-ballistic phonon transport.

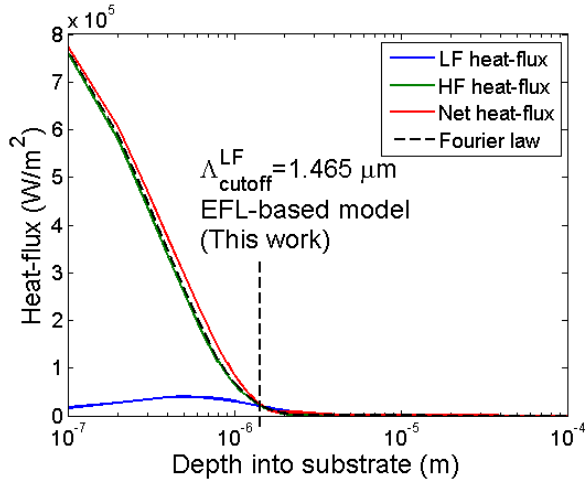

Fig. 4: $\Lambda^{LF}$ in Eq. (2) is varied iteratively until the LF mode heat-flux equals the HF-mode heat-flux, and the resulting value (1.465 microns, as shown) is taken as the actual cutoff. Then we find that only 6% of the net heat-flux at a depth of 376 nm (see Fig. 1) comes from LF modes. Also, HF mode heat-flux agrees well with "effective" Fourier law (dashed line).

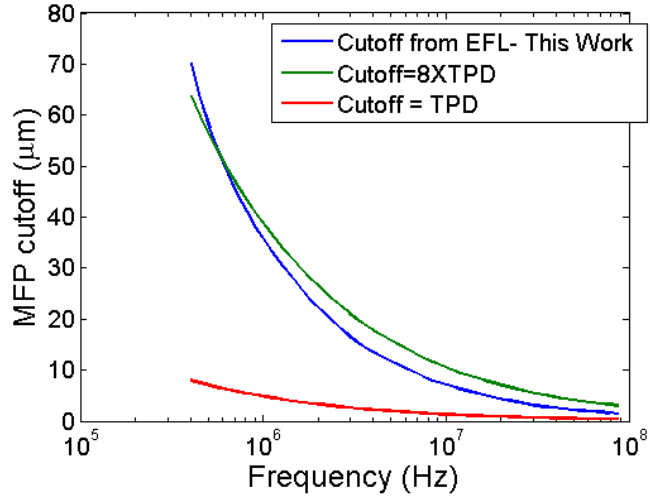

Fig. 5: The cutoff MFP is 4X TBD at high frequencies, and rises to about 9X TPD at low frequencies. Green curve shows fit of EFL-based cut-off with 8X TPD; error is large at high frequencies (by a factor of 2).

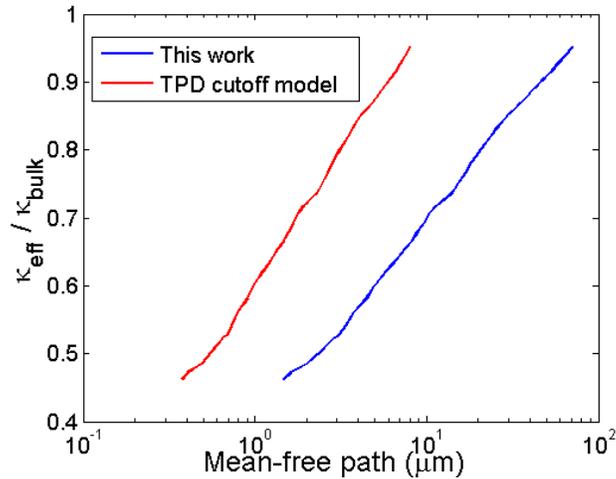

Fig. 6: The mean-free path accumulation function; red solid curve is after Regner *et al.* [6]. Our values (blue solid curve) are shifted considerably to the right due to contributions from quasi-ballistic LF-mode phonons.

## 5. Conclusions

We have utilized the enhanced Fourier law proposed earlier by one of the authors to develop a novel framework for evaluating the mean-free path accumulation function for effective thermal conductivity from Fourier domain thermoreflectance experiments. We have seen that cutting off the quasi-ballistic modes at the thermal penetration depth at each frequency neglects their important contribution to the heat-flux, especially at high frequencies. Our more rigorous criterion has resulted in mean-free path cutoffs much larger than the thermal penetration depth. This has changed drastically the accumulation function, a highly important quantity for thermal transport theory and measurements. The current limitation of our framework is that it is in one dimension, while practical FDTR experiments require a two-dimensional enhanced Fourier law for their analyses due to small pump-beam spot sizes (3-15 microns).

Appendix A

We derive Eqs. (1)-(4) from the model assumptions stated in Sec. 2 (Theoretical Model). The derivation follows [15] closely.

We denote the distribution function for LF modes as $g(x, \mathbf{k})$, where all spatial variation is assumed to be along the x-direction, and $\mathbf{k}$ is the phonon mode wave-vector of magnitude $k$ and making an angle $\theta$ with the x-axis. We expand the distribution function in terms of spherical harmonics $P_l(cos\theta)$. Since spherical harmonics are eigen-functions of the Legendre differential equation which is a Sturm-Liouville equation, they form an orthogonal basis for expanding angle-dependent azimuthally symmetric functions [26]:

$$g(x, \mathbf{k}) = \sum_{l=0}^{\infty} g_l(x, k) P_l(cos\theta) \tag{5}$$

The steady-state linearized BTE for the LF modes is given by

$$v cos\theta \frac{\partial g(x,\mathbf{k})}{\partial x} = -\frac{g(x,\mathbf{k}) - f_{Eq}(x,k,T)}{\tau} \tag{6}$$

We begin with the observation that, owing to the orthogonality of the spherical harmonics, the x-component of the LF heat-flux is determined solely by the first spherical harmonic $g_1$:

$$q^{LF} = 2\pi \sum_k \int_{\theta=0}^{\pi} \hbar\omega g(x,\mathbf{k}) v cos\theta sin\theta d\theta = \frac{4\pi}{3} \sum_k \hbar\omega v g_1 \tag{7}$$

Therefore we seek a differential equation for $g_1$. Here and henceforth, $\sum_k I(k)$ is shorthand for $\frac{1}{(2\pi)^3} \int dk I(k) k^2$ where $I(k)$ is any function of $k$, and the integral is over all LF mode wave-vector magnitudes. Substituting Eq. (1) into the Boltzmann transport equation, Eq. (2), multiplying successively by $P_{l'}(cos\theta) sin\theta$ for $l' = 0,1,2,...$ and integrating over $\theta$, we arrive at a hierarchy of coupled equations for the $g_l$s, [27], the first three of which are

$$\frac{1}{3} v \frac{\partial g_1}{\partial x} + \frac{g_0 - f_{Eq}(T)}{\tau} = 0 \tag{8a}$$

$$\frac{2}{5} v \frac{\partial g_2}{\partial x} + v \frac{\partial g_0}{\partial x} + \frac{g_1}{\tau} = 0 \tag{8b}$$

$$\frac{3}{7} v \frac{\partial g_3}{\partial x} + \frac{2}{3} v \frac{\partial g_1}{\partial x} + \frac{g_2}{\tau} = 0 \tag{8c}$$

We truncate the hierarchy by setting $g_3 = 0$; other truncations are possible [27]. Substituting Eq. (8c) into Eq. (8b) to eliminate $g_2$, and the result into Eq. (8a) to eliminate $g_0$, we arrive at an equation expressed solely in terms of $g_1$:

$$-\frac{3}{5}(v\tau)^2 \frac{\partial^2 g_1}{\partial x^2} + v\tau \frac{\partial f_{Eq}(T)}{\partial x} + g_1 = 0 \tag{9}$$

$f_{Eq}(T)$ depends on x only through T, enabling the replacement $\frac{\partial f_{Eq}(T)}{\partial x} = \frac{\partial f_{Eq}(T)}{\partial T} \frac{dT}{dx}$. Multiplying Eq. (9) by $\frac{4\pi}{3} \hbar\omega v$ and summing over all $k$,

$$-\frac{3}{5}(v\tau)^2 \frac{\partial^2 q^{LF}}{\partial x^2} + \frac{1}{3} C^{LF} v^2 \tau \frac{\partial T}{\partial x} + q^{LF} = 0 \tag{10}$$

Here $C^{LF} = 4\pi \frac{\partial}{\partial T} \sum_k \hbar\omega f_{Eq}(T)$ is the volumetric heat-capacity of the LF modes, an equilibrium thermodynamic property.

Defining $\kappa^{LF} = \frac{1}{3} C^{LF} v^2 \tau$ as the thermal conductivity of the LF modes, in analogy with the expression of the kinetic theory [24], and $\Lambda^{LF} = v\tau$, the MFP of low-frequency phonons, Eq. (10) reads

$$q^{LF} = \frac{3}{5}(\Lambda^{LF})^2 \frac{\partial^2 q^{LF}}{\partial x^2} - \kappa^{LF} \frac{\partial T}{\partial x} \tag{11}$$

A similar analysis is carried out for the high-frequency distribution,

$$h(x, \mathbf{k}) = \sum_{l=0}^{\infty} h_l(x, k) P_l(\cos\theta) \tag{12}$$

with two distinctions. First, we truncate the SHE at the first order, $h(x, \mathbf{k}) = h_0 + h_1 \cos\theta$. Second, we assume quasi-Bose statistics for the symmetric part of this distribution, that is, $h_0 = f_{Eq}(T)$. The result of this analysis is

$$q^{HF} = -\kappa^{HF} \frac{\partial T}{\partial x} \tag{13}$$

Adding Eq. (11) and Eq. (13), and defining the total thermal conductivity $\kappa = \kappa^{HF} + \kappa^{LF}$,

$$q = q^{LF} + q^{HF} = \frac{3}{5}(\Lambda^{LF})^2 \frac{\partial^2 q^{LF}}{\partial x^2} - \kappa \frac{\partial T}{\partial x} \tag{14}$$

Adding and subtracting $\frac{3}{5}(\Lambda^{LF})^2 \frac{\partial^2 q^{HF}}{\partial x^2}$ where $q^{HF}$ is known from Eq. (13), we finally arrive at an equation for the heat-flux:

$$q = \frac{3}{5}(\Lambda^{LF})^2 \frac{\partial^2 q}{\partial x^2} + \frac{3}{5}\kappa^{HF}(\Lambda^{LF})^2 \frac{\partial^3 T}{\partial x^3} - \kappa \frac{\partial T}{\partial x} \tag{15}$$

This is to be combined with energy conservation for the total energy density $E$, in the presence of an external source of heat $S^{HF}(x, t)$ that couples only to the HF modes

$$\frac{\partial q}{\partial x} = -\frac{\partial E}{\partial t} + S^{HF}(x, t) \tag{16}$$

We now make the first approximation beyond the truncation of the SHEs, namely that $E \sim CT$ where the total heat capacity $C$ equals $C^{HF}$, the heat capacity of the HF modes only. In other words, we assume that the heat-capacity of the LF modes satisfies $C^{LF} \ll C^{HF}$. This assumption is usually included as part of the two-fluid model [16] as described in Sec. 2, and its plausibility may be roughly justified by noting that in a simple Debye model, the density-of-state, and hence the heat capacity per unit energy, varies as the square of the frequency [24]. The thermal conductivities of the two channels however may still be comparable because of their additional dependence on the respective MFPs. Although Eq. (16) is a statement of overall energy conservation, we exclude the coupling of external heat sources to the LF modes in order to maintain consistency with the omission of source terms from Eq. (6) for the LF modes.

Thus we arrive at our result, the enhanced Fourier law, which comprises the set of two coupled equations:

$$\frac{\partial q}{\partial x} = -C \frac{\partial T}{\partial t} + S^{HF}(x, t) \tag{17a}$$

$$q = \frac{3}{5}(\Lambda^{LF})^2 \frac{\partial^2 q}{\partial x^2} + \frac{3}{5}\kappa^{HF}(\Lambda^{LF})^2 \frac{\partial^3 T}{\partial x^3} - \kappa \frac{\partial T}{\partial x} \qquad (17b)$$

The basic approximation of truncation of the spherical harmonic expansion at the second order needs more justification for quasi-ballistic phonon modes. For this, we refer the reader to [25], where a generalized form of the enhanced Fourier law is compared against a closed form solution of the BTE for the transient grating experiment, and nearly perfect agreement is obtained over the entire range of values of the experimental length-scale.

Acknowledgments:

We are grateful to Dr. Alexei Maznev (Massachusetts Institute of Technology) for several helpful discussions. We also wish to thank Professor Jonathan Malen (Carnegie Mellon University) for giving us access to raw data from the paper, Regner *et al.*, which was used to construct Figs. 1 and 6. This work was funded by the National Science Foundation (NSF) under contract number CMMI-1363207.